\documentclass{aa}
\usepackage{times,float}
\usepackage{epsfig}
\usepackage{graphics}
\usepackage{amssymb}
\usepackage{curves}
\usepackage{eepic}
\usepackage{epic}

\begin{document}

\title{A multi-colour study of the dark GRB~000210 host galaxy and its
environment  \thanks{Based on observations  collected at  the European
Southern  Observatory, in  La  Silla and  Paranal  (Chile), ESO  Large
Programmes    165.H-0464(C),    165.H-0464(E),    165.H-0464(I)    and
265.D-5742(C), granted to the  GRACE Team.  Based on observations made
with the Danish 1.54 m telescope, in La Silla (Chile). }}

\titlerunning{The host galaxy  of GRB~000210}
{\small
\author{
        J. Gorosabel \inst{1,2,3}
   \and L. Christensen \inst{4,5}
   \and J. Hjorth \inst{4}
   \and J.U. Fynbo \inst{6,4}
   \and H. Pedersen \inst{4}
   \and B.L. Jensen \inst{4}
   \and M.I. Andersen \inst{5}
   \and N. Lund \inst{1}
   \and A.O. Jaunsen \inst{7}
   \and J.M. Castro Cer\'on \inst{8}
   \and A.J. Castro-Tirado \inst{2}
   \and A. Fruchter \inst{9}
   \and J. Greiner \inst{5,10}
   \and E. Pian \inst{11}
   \and P. M. Vreeswijk \inst{7}
   \and I. Burud \inst{9}
   \and F. Frontera \inst{12,13}
   \and L. Kaper \inst{14}
   \and S. Klose \inst{15}
   \and C. Kouveliotou \inst{16}
   \and N. Masetti \inst{13}
   \and E. Palazzi \inst{13}
   \and J. Rhoads \inst{9}
   \and E. Rol \inst{14}
   \and I. Salamanca \inst{14}
   \and N. Tanvir \inst{17}
   \and R.A.M.J. Wijers \inst{14}
   \and E. van den Heuvel \inst{14}
}
}

\institute{ 
           Danish Space Research Institute,
           Juliane Maries Vej 30, DK--2100 Copenhagen \O, Denmark;
           {\tt jgu@dsri.dk, nl@dsri.dk}
           \and
           Instituto de Astrof\'{\i}sica de Andaluc\'{\i}a (IAA-CSIC),
           P.O. Box 03004, E-18080 Granada, Spain; {\tt jgu@iaa.es, ajct@iaa.es}
            \and
            Laboratorio de Astrof\'{\i}sica Espacial y F\'{\i}sica
            Fundamental
            (LAEFF-INTA), P.O. Box 50727, E-28080, Madrid, Spain;
            {\tt jgu@laeff.esa.es}
           \and
           Astronomical Observatory,
           University of Copenhagen,
           Juliane Maries Vej 30, DK--2100 Copenhagen \O, Denmark;
           {\tt lise@astro.ku.dk, jens@astro.ku.dk, holger@astro.ku.dk, brian\_j@astro.ku.dk}
           \and
           Astrophysikalisches Institut, 14482 Potsdam, Germany;
           {\tt lchristensen@aip.de, mandersen@aip.de, jgreiner@aip.de}
           \and
           Department of Physics and Astronomy, University of \AA rhus,  Ny
           Munkegade, 8000 \AA rhus C, Denmark; {\tt jfynbo@phys.au.dk}
           \and
           European Southern Observatory, Casilla 19001, Santiago 19,
           Chile; {\tt ajaunsen@eso.org,  pvreeswi@eso.org}
           \and
           Real   Instituto  y Observatorio   de   la Armada,  Secci\'on de
           Astronom\'\i a, 11.110 San  Fernando-Naval (C\'adiz), Spain; \\
          {\tt  josemari@alumni.nd.edu}
           \and
           Space Telescope Science Institute,
           3700 San Martin Drive, Baltimore, MD 21218, USA; \linebreak
           {\tt fruchter@stsci.edu, burud@stsci.edu, rhoads@stsci.edu}
           \and
           Max-Planck-Institut f\"ur extraterrestrische Physik, 85741 Garching, Germany;
           {\tt jcg@mpe.mpg.de}
           \and
           Osservatorio Astronomico di Trieste, Via G.B. Tiepolo 11, 34131,
           Trieste, Italy; {\tt pian@tesre.bo.cnr.it}
           \and
            Dipartimento di Fisica, Universit\`a di Ferrara,
            Via Paradiso 12, I-44100 Ferrara, Italy; {\tt frontera@fe.infn.it}
           \and
           Istituto Tecnologie e Studio Radiazioni Extraterrestri,
           CNR, Via Gobetti 101, 40129 Bologna, Italy; \\
           {\tt filippo@bo.iasf.cnr.it, masetti@bo.iasf.cnr.it, eliana@bo.iasf.cnr.it}
           \and
           University of Amsterdam,
           Kruislaan 403, 1098 SJ Amsterdam, The Netherlands;
           {\tt lexk@science.uva.nl,evert@science.uva.nl,isabel@science.uva.nl, rwijers@science.uva.nl, edvdh@science.uva.nl}
           \and                      
           Th\"uringer Landessternwarte Tautenburg, D-07778 Tautenburg,
           Germany;        {\tt  klose@tls-tautenburg.de}
           \and
           NASA MSFC, SD-50, Huntsville, AL 35812,
           USA; {\tt kouveliotou@eagles.msfc.nasa.gov}
           \and
           Department of   Physical Sciences, University  of Hertfordshire,
           College     Lane,    Hatfield,   Herts AL10     9AB,  UK;   {\tt
             nrt@star.herts.ac.uk}
           }

\offprints{ \hbox{J. Gorosabel, e-mail:{\tt  jgu@dsri.dk}}}

\date{Received  / Accepted }


\abstract{We  present UBVRIZJsHKs  broad band  photometry of  the host
  galaxy  of the  dark gamma-ray  burst  (GRB) of  February 10,  2000.
  These observations represent the most exhaustive photometry given to
  date of any GRB host galaxy.  A grid of spectral templates have been
  fitted to the  Spectral Energy Distribution (SED) of  the host.  The
  derived  photometric redshift is  $z=0.842^{+0.054}_{-0.042}$, which
  is   in  excellent   agreement  with   the   spectroscopic  redshift
  ($z=0.8463\pm  0.0002$)  proposed by  Piro  et al.\  (\cite{Piro02})
  based on  a single emission  line.  Furthermore, we  have determined
  the  photometric  redshift  of  all  the  galaxies  in  an  area  of
  $6^{\prime} \times  6^{\prime}$ around the host galaxy,  in order to
  check for their overdensity in the environment of the host.  We find
  that the GRB~000210 host galaxy is a subluminous galaxy ($L \sim 0.5
  \pm  0.2  L^{\star}$),  with   no  companions  above  our  detection
  threshold  of $0.18  \pm 0.06  L^{\star}$.  Based  on  the restframe
  ultraviolet flux  a star formation  rate of $2.1 \pm  0.2 M_{\odot}$
  yr$^{-1}$ is estimated.   The best fit to the SED  is obtained for a
  starburst template with an  age of $0.181^{+0.037}_{-0.026}$ Gyr and
  a  very  low  extinction  ($A_{\rm  V} \sim  0$).   We  discuss  the
  implications of the inferred low value of $A_{\rm V}$ and the age of
  the  dominant  stellar  population  for  the non  detection  of  the
  GRB~000210  optical  afterglow.   }

\maketitle

\section{Introduction}

The origin of cosmological Gamma-Ray  Bursts (GRBs) remains one of the
great    mysteries   of    modern   astronomy    (van    Paradijs   et
al. \cite{vanP00}).  Over the past half decade many advances have been
made in  understanding the nature  of the bursts and  their afterglows
throughout the electromagnetic spectrum.   There are at present mainly
two sets  of models for  GRBs.  One set  of models predicts  that GRBs
occur  when two  collapsed objects  (such  as black  holes or  neutron
stars)  merge  (Eichler  et  al.  \cite{Eich89};  Mochkovitch  et  al.
\cite{Moch93}).  The time-scale for binary compact objects to merge is
large  ($\gtrapprox$ 1  Gyr), so  GRBs  can occur  after massive  star
formation  has ended  in  a galaxy.   The  other major  set of  models
predicts  that GRBs  are associated  with the  death of  massive stars
(supernovae   or  hypernovae)   (Woosley   \cite{Woos93};  Paczy\'nski
\cite{Pacz98}; MacFadyen \& Woosley \cite{MacF99}).  In this case GRBs
will  coincide with  the  epoch of  star  formation in  the host.   By
determining the Spectral Energy  Distribution (SED) and star formation
rate (SFR) of a sample of GRB host galaxies we can distinguish between
these two  families of GRB  progenitor models (see also  Belczynski et
al. \cite{Belc02}).  Substantial insight has already been gained about
the galaxies that the bursts occur in.  Radio, optical and/or infrared
afterglows have been  observed for $\sim$40 GRBs, and  the majority of
these coincide with starforming galaxies.

As GRB host galaxies tend to be faint (R$ > 23$) spectroscopic studies
of  the SED  are  only reachable  with  8--10 m  class telescopes.   A
cheaper  and  elegant  alternative   to  spectroscopy  is  to  extract
information  on   the  properties  of  the  host   galaxies  based  on
multicolour  broad band imaging.   By determining  the colours  of GRB
host galaxies  we can derive or  constrain the age  of the predominant
stellar population  as well as the  extinction. As part  of the global
fit, the photometric  redshift of the host galaxies  can be derived if
the redshift  is not known in advance  from spectroscopic observations
of the afterglow and/or the host galaxy.  Additional advantages of the
multicolour photometric  studies compared to  spectroscopic techniques
are  their   simplicity  and  their   multi-object  feasibility.   The
photometric technique  allows the determination of the  colours of all
objects  in the  field  down to  the  imaging flux  limit, thereby  in
principle  permitting the study  of the  host galaxy  environment. The
precision of  the photometric redshift estimate (which  depends on the
photometric accuracy,  the spectral coverage and the  number of bands)
is  evidently not  as accurate  as the  spectroscopic one,  but  it is
sufficient for a first order study of host galaxy environments.

So  far it has  been possible  to detect  optical afterglows  for only
about 30\% of localised GRBs (Fynbo et al.  \cite {Fynb01}; Lazzati et
al.  \cite{Lazz02}).  It is important  to understand the nature of the
remaining (rather ill-termed) so-called dark  GRBs if we wish to get a
complete understanding on GRB  selected galaxies and thereby constrain
the  GRB  progenitors as  well  as  the  distribution of  cosmic  star
formation    over    different    modes   (e.g.     Ramirez-Ruiz    et
al. \cite{Rami02};  Venemans \&  Blain \cite{Vene01}).  GRB~000210 is
currently one  of only few  systems that allow  a detailed study  of a
galaxy  hosting   a  dark  GRB.   The  burst   exhibited  the  highest
$\gamma$-ray peak flux  among the 54 GRBs localized  during the entire
BeppoSAX operation,  from 1996 to  2002 (Piro et  al.  \cite{Piro02}).
However, no  optical afterglow  (OA) was detected  in spite of  a deep
search (R$> 23.5$) carried out  $\sim16$ hrs after the gamma-ray event
(Gorosabel et al.  \cite{Goro00a}).  X-ray observations performed with
the Chandra X-ray  telescope 21 hrs after the  GRB localised the X-ray
afterglow of  the burst to  an accuracy of $2^{\prime  \prime}$, later
improved by Piro et al.   (\cite{Piro02}) to a $0\farcs6$ radius error
circle.   The  optical search  revealed  an  extended constant  source
coincident with the X-ray afterglow which was proposed as the GRB host
galaxy (Gorosabel  et al.  \cite{Goro00b}).  In addition,  Piro et al.
(\cite{Piro02}) have  reported the detection  of a radio  transient at
8.5 GHz spatially  coincident with the X-ray afterglow.   Based on the
detection of a single host galaxy spectral line, interpreted to be due
to [\ion{O}{II}], Piro et  al.  (\cite{Piro02}) proposed a redshift of
$z=0.8463 \pm  0.0002$.  Recently  Berger et al.   (\cite{Berg02}) and
Barnard  et  al.   (\cite{Barn02})  have reported  $\sim$2.5  $\sigma$
detections  of sub-mm  emission  towards the  position  of GRB~000210
interpreted as  emission from the  host galaxy and hence  suggesting a
SFR of several hundred $M_{\odot}$ yr$^{-1}$.

\begin{table*}
\begin{center}
\caption{Chronologically ordered optical  and NIR observations carried
out for the GRB~000210 host galaxy.}
\begin{tabular}{lcccccr}
Telescope     &  Filter & Date UT &T$_{\rm exp}$&  Seeing  & Limiting magnitude \\
(+Instrument) &         &         &  (s)       &          &    ($4\sigma$)     \\
\hline
 8.2VLT (+FORS1)  &  R  &  25.237--25.240/10/00&           300 & 0.70& 25.4$^{\star\star}$&\\
3.58NTT (+SOFI)   &  H  &  02.251--02.410/09/01& 182$\times$60 & 0.90& 22.8\\
 3.6ESO (+EFOSC2) &  V  &  13.219--13.253/09/01&   4$\times$600& 1.75& 25.4\\
 3.6ESO (+EFOSC2) &  I  &  13.256--13.278/09/01&   3$\times$600& 1.45& 23.1\\
 3.6ESO (+EFOSC2) &  B  &  13.280--13.302/09/01&   3$\times$600& 1.70& 25.6\\
 3.6ESO (+EFOSC2) &  U  &  13.304--13.348/09/01&   6$\times$600& 1.55& 24.7\\
 8.2VLT (+ISAAC)  &  Ks &  21.159--21.193/09/01&  30$\times$60 & 0.45& 22.2\\
 8.2VLT (+ISAAC)  &  Js &  21.194--21.218/09/01&  15$\times$120& 0.60& 24.1$^{\dag}$ \\
 8.2VLT (+ISAAC)  &  Js &  23.193--23.218/09/01&  15$\times$120& 0.75& 24.1$^{\dag}$ \\
 1.54D  (+DFOSC)  &  Z  &  19.090--19.254/12/01&  14$\times$600& 1.10& 22.9$^{\star}$ \\
 1.54D  (+DFOSC)  &  Z  &  20.042--20.394/12/01&  21$\times$600& 1.15& 22.9$^{\star}$ \\
\hline
\multicolumn{7}{l}{$\star\star$     Published in Piro et al. (\cite{Piro02}).}\\
\multicolumn{7}{l}{$\dag$     The images were coadded resulting in just a single Js-band magnitude.}\\
\multicolumn{7}{l}{$\star$    The images were coadded resulting in just a single Z-band magnitude.}\\
\hline             
\label{table1}
\end{tabular}
\end{center}
\end{table*}

In this paper  we present the most intensive  multi-colour host galaxy
imaging performed to date.  The  host galaxies SED studies to date had
a limited number of bands (Sokolov et al.  \cite{Soko01}; Chary et al.
\cite{Char02})    and   no   photometric    redshift   determinations.
Throughout, the  assumed cosmology  will be $\Omega_{\Lambda}  = 0.7$,
$\Omega_{M} =  0.3$ and  $H_0= 65$ km  s$^{-1}$ Mpc$^{-1}$  (except in
Sect.  \ref{subluminous} where the  host galaxy luminosity is rescaled
to  the  cosmology used  by  Lilly  et  al.  \cite{Lill95}).   At  the
proposed spectroscopic  redshift ($z=0.8463$),  the look back  time is
7.59 Gyr  (52.4\% of the present  age) and the  luminosity distance is
$1.79 \times 10^{28}$ cm.  The  physical transverse size of one arcsec
at $z=0.8463$ corresponds to 8.24 kpc.

\section{Observations and photometry}
\label{Observations}

We have used  a number of optical/near-IR (NIR)  resources in order to
compose a well sampled SED (see Table~\ref{table1}). UBVI observations
were carried out  with the 3.6-m ESO telescope  (3.6ESO) equipped with
EFOSC2, covering a field of  view (FOV) of $5\farcm5 \times 5\farcm5$.
These  observations  were  carried  out in  2$\times$2  binning  mode,
providing a pixel scale  of $0\farcs32$/pix.  R-band measurements were
obtained  with the  UT1 of  the  8.2-m Very  Large Telescope  (8.2VLT)
equipped with FORS1 and are published in Piro et al.  (\cite{Piro02}).
The Z-band observations were carried out during two consecutive nights
with the  1.54-m Danish Telescope  (1.54D) equipped with  DFOSC, which
provides a FOV of $13\farcm 7  \times 13\farcm 7$ and a pixel scale of
$0\farcs39$/pix.

\begin{table*}
\begin{center}
\caption{Magnitudes  of the  host in  the UBVRIZJsHKs  bands.  Several
characteristics  of  the  filters  are  displayed:  filter  name  (1),
effective  wavelength (2) and  bandpass width  (3). The  fourth column
shows the  measured magnitudes (in  the Vega system and  not corrected
from  Galactic reddening).  To  facilitate the  calculation of  the AB
magnitudes, and consequently the flux  densities for each band, the AB
offsets are provided in the fifth column. }
\begin{tabular}{@{}lcccc@{}}
\hline
Filter              & Effective       & Bandpass    &Magnitude &ABoff\\
name                & wavelength (\AA)& width (\AA) &          &     \\
\hline                                                                      
 U (ESO\#640)       &   3718.8 &  172.9 & 23.54$\pm$0.13 &$ 0.73$\\
 B (ESO\#639)       &   4372.6 &  701.4 & 24.40$\pm$0.13 &$-0.07$\\
 V (ESO\#641)       &   5563.9 &  856.4 & 24.22$\pm$0.08 &$ 0.04$\\
 R (ESO R\_BESSEL+36)&  6608.5 & 1300.3 & 23.46$\pm$0.10$^{\dag}$&$ 0.23$\\
 I (ESO\#705)       &   7950.2 &  844.0 & 22.49$\pm$0.12 &$ 0.45$\\
 Z (ESO\#462)       &   9477.4 &  985.1 & 22.83$\pm$0.28 &$ 0.56$\\
Js (ISAAC)          &  12498.9 &  957.8 & 21.98$\pm$0.10 &$ 0.94$\\
 H (SOFI)           &  16519.6 & 1732.3 & 21.51$\pm$0.23 &$ 1.41$\\
Ks (ISAAC)          &  21638.0 & 1637.9 & 20.94$\pm$0.14 &$ 1.87$\\
\hline
\multicolumn{5}{l}{$\dag$ Published in Piro et al. (\cite{Piro02}).}\\
\hline
\label{table2}
\end{tabular}
\end{center}
\end{table*}

The H-band  observations were acquired with the  3.58-m New Technology
Telescope (3.58NTT) using SOFI in the large FOV mode, which provides a
FOV  of   $4\farcm  9  \times  4\farcm   9$  and  a   pixel  scale  of
$0\farcs292$/pix.  The  Js and Ks-band  observations are based  on the
UT1 of the  8.2VLT equipped with ISAAC, allowing us to  cover a FOV of
$2\farcm 5 \times  2\farcm 5$ with a pixel  scale of $0\farcs148$/pix.
In Table \ref{table1} we provide  the observing log of our optical and
NIR observations.

\begin{figure*}[t]
\begin{center}
     {\includegraphics[width=\hsize]{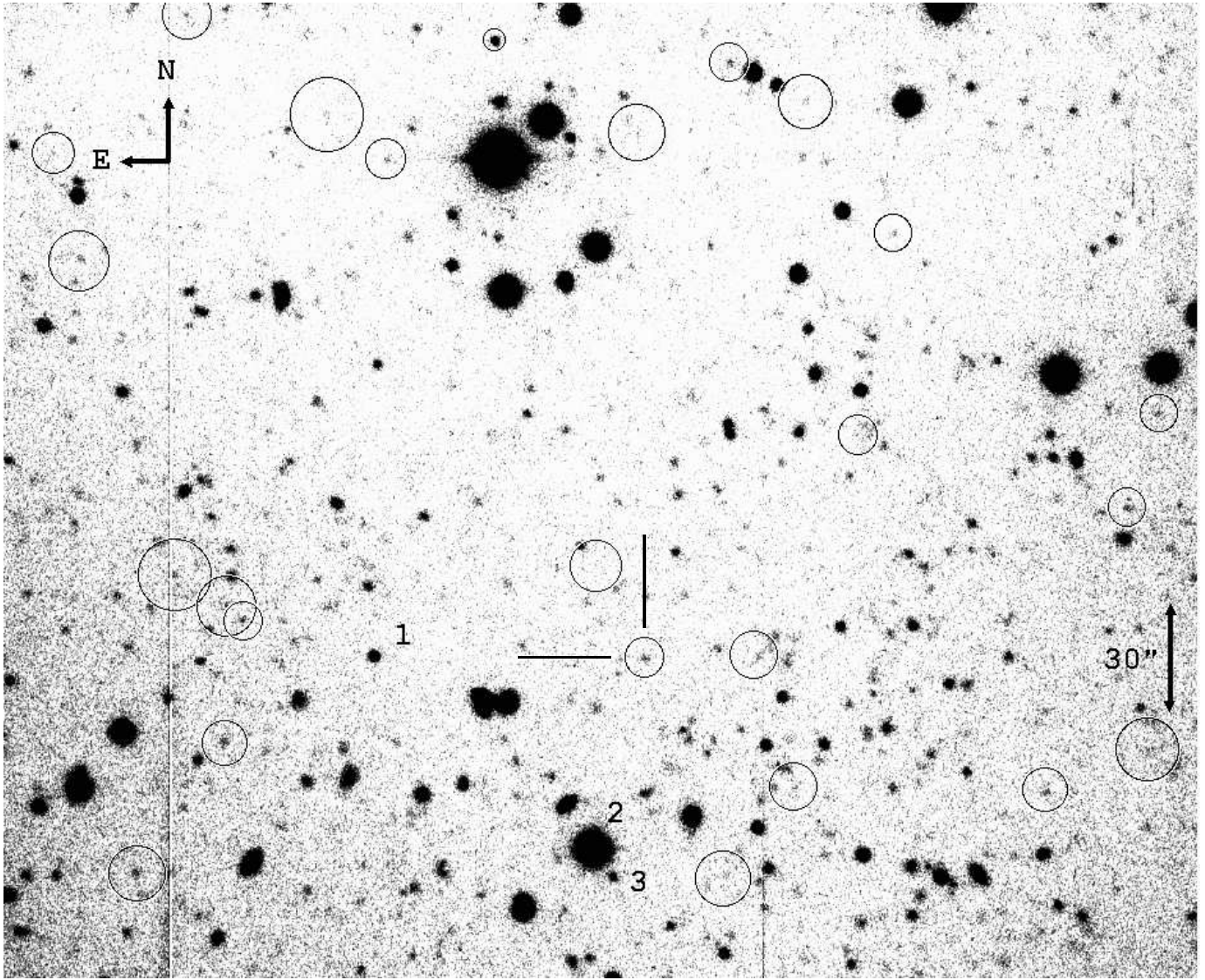}}
       \Thicklines
       \drawline(-235.5,150)(-235.5,190)
       \drawline(-290,137)(-250,137)
\caption{\label{3.6ESOV}  The image  shows the  co-added  V-band image
taken at the 3.6ESO  telescope at 13.219--13.253/09/01 UT. The objects
contained in the  circles are the ones with  redshifts consistent with
$0.746  <  z  < 0.946$.   As  it  can  be  seen  there is  no  obvious
concentration of these galaxies around the host.  The circle radius is
proportional to  $1/|M_B|$, so the  fainter the galaxy the  larger the
circle.  The host galaxy is  indicated by the tick marks.  The numbers
label the secondary NIR standards shown in Table~\ref{standards}.  The
FOV   covered  by   the  image   corresponds  to   $5^{\prime}  \times
5^{\prime}$.}
\end{center}
\end{figure*}

Given  that  every  extended  source shows  a  different  photodensity
profile  (or FWHM),  an unique  fixed Aperture  Photometry  (or static
aperture photometry, AP hereafter) would yield unsatisfactory results.
On the  other hand, Isophotal  Photometry (IP) would also  not provide
optimum photometry, since performing IP we would not consider the same
fraction of each galaxy in the different bands due to colour-dependent
morphologies and  seeing.  To solve this problem  the total integrated
photometry   given  by   SExtractor  was   used  (Bertin   \&  Arnouts
\cite{Bert96}).   For each  object  SExtractor performs  two types  of
total integrated photometry:  the Adaptative Aperture Photometry (AAP)
and  the  Corrected  Isophotal  Photometry  (CIP).  The  AAP  and  CIP
supersede the values given by the AP and IP, respectively, applying to
them an aperture correction.  For each object SExtractor considers the
photometry output  given by  the AAP, except  if a neighbour  is found
biasing the flux  by more than 10\%.  If this  is the case, SExtractor
chooses  the   value  given  by   the  CIP  (see  Bertin   \&  Arnouts
\cite{Bert96}  for details).  The  host galaxy  of GRB~000210  is well
isolated  and  hence  its  the  photometry  is  not  affected  by  any
neighbours.

The  UBVRIZJsHKs-band magnitudes  of the  host  can be  seen in  Table
\ref{table2}.  The  UBVRI-band calibration  is based on  the secondary
standards  given in  Table 2  of  Piro et  al.  (\cite{Piro02}).   The
JHKs-band  calibration  was performed  observing  the standard  fields
sj9105  and   sj9172  (Persson  et  al.    \cite{Pers98})  at  several
airmasses.   The derived NIR  secondary standards  are given  in Table
\ref{standards}  and  displayed  in  Fig.~\ref{3.6ESOV}.   The  Z-band
calibration was carried out observing the spectro-photometric standard
stars LTT2415 and LTT1788 (Hamuy et al.  \cite{Hamu94}) with the 1.54D
at  an airmass  similar to  that of  the GRB  field.  The  host galaxy
BVRI-band  magnitudes reported  by  Piro et  al.  (\cite{Piro02})  are
consistent with our magnitudes displayed in Table \ref{table2}.

\begin{table*}[t]
\caption{\label{standards} NIR secondary standards in the GRB~000210 field.}
\begin{center}
\begin{tabular}{lccccc}
\hline
 Name &$\alpha_{2000}$&$\delta_{2000}$&  Js &       H       &       Ks      \\
\hline
 1 & 1:59:21.51 & -40:39:33.4 & 17.94$\pm$0.03  & 17.19$\pm$0.03  & 17.00$\pm$0.03 \\
 2 & 1:59:16.72 & -40:40:20.3 & 16.77$\pm$0.07  & 16.51$\pm$0.03  & 16.59$\pm$0.04 \\
 3 & 1:59:16.27 & -40:40:27.2 & 18.35$\pm$0.08  & 17.69$\pm$0.04  & 17.46$\pm$0.05 \\
\hline
\hline
\end{tabular}
\end{center}
\end{table*}

In  order to  derive the  corresponding effective  wavelengths  and AB
offsets  we   convolved  each  filter  transmission   curve  with  the
corresponding CCD  efficiency curve (see Table  \ref{table2}).  The AB
offset  is defined  as  ABoff$= m_{\rm  AB}  - m$,  where  $m$ is  the
magnitude in the Vega system and  $m_{\rm AB}$ is the magnitude in the
AB system (given  by $m_{\rm AB} = -2.5 \times  \log f_{\nu} - 48.60$,
being $f_{\nu}$ the flux density in erg s$^{-1}$ cm$^{-2}$ Hz$^{-1}$).

The AB offsets of the nine bands have been derived convolving the Vega
spectrum taken  from the  GISSEL98 (Bruzual \&  Charlot \cite{Bruz93})
library ($\alpha$  Lyrae $m=0$  in all bands  by definition)  with our
UBVRIZJsHKs-band filters and  the corresponding CCD efficiency curves.
The  derived  AB  offsets  (displayed  in the  last  column  of  Table
\ref{table2})  are  similar  to  the  ones  reported  by  Fukugita  et
al. (\cite{Fuku95}).

\begin{table*}
\begin{center}
\caption{The table displays the parameters of the best host galaxy SED
fit when  several IMFs,  indicated in the  first column,  are adopted.
The  rest of  the columns  display the  inferred parameters  under the
assumed IMF.   The second column  provides the confidence of  the best
fit (given  by $\chi^{2}/$dof).   The derived photometric  redshift is
displayed in  the third  column (and the  corresponding 68\%  and 99\%
percentile  errors).  In  the fourth  and fifth  columns  the template
family of  the best fitted SED  and the age of  the stellar population
are given.   The sixth column displays  the derived value  of the host
galaxy  extinction  $A_{\rm  V}$.   The seventh  column  displays  the
derived  rest frame absolute  B-band magnitude,  $M_B$.  The  last two
columns give the Luminosity of  the host in units of $L^{\star}$, when
the luminosity functions of Schechter (\cite{Sche76}) and Lilly et al.
(\cite{Lill95}) are used (see  Sect.  \ref{subluminous} for a detailed
discussion).  The extinction law has  been fixed to follow Calzetti et
al.   (\cite{Calz00}) (the  effect of  the adopted  extinction  law is
discussed in Sect.   \ref{effect}). As shown in the  first two rows of
the table, the  resolution of our template grid is not  able to make a
distinction between  most of the properties (Age,  $A_{\rm V}$, $M_B$,
$L/L^{\star}$) derived for the  Sa55 and MiSc79 IMFs.  The photometric
redshifts derived for  the three IMFs are consistent,  within the 99\%
percentile  error range,  with the  spectroscopic  redshift.  However,
within the 68\% precentile ($\sim 1\sigma$) error range, only the Sa55
and MiSc79 IMFs are consistent, Sc86 is not.}

\begin{tabular}{@{}lcccccccc@{}}
\hline
 IMF        & $\chi^{2}/$dof& Photometric redshift & Template &  Age  & $A_{\rm V}$& $M_B$&$L/L^{\star}$&$L/L^{\star}$\\

                                &           &$z^{+ p68\%, p99\%}_{- p68\%, p99\%}$&& (Gyr)&            &      &    &\\
                                &           &                          &          &       &            &      &    &\\
\hline                                                                                                              
 Salpeter (\cite{Salp55})       & $1.096$   &$0.842^{+0.054, 0.158}_{-0.042, 0.279}$& Stb &0.181& 0.00 &-20.16&0.67&0.35\\
                                &           &                          &          &       &            &      &    &\\
 Miller \& Scalo (\cite{Mill79})& $1.046$   &$0.836^{+0.087, 0.140}_{-0.053, 0.244}$& Stb &0.181& 0.00 &-20.16&0.67&0.35\\
                                &           &                          &          &      &             &      &    &\\
 Scalo (\cite{Scal86})          & $0.903$   &$0.757^{+0.067, 0.219}_{-0.044, 0.132}$& S0  &1.015& 0.00 &-19.90&0.52&0.27\\
                                &           &                          &          &      &             &      &    &\\
\hline
\label{table3}
\end{tabular}
\end{center}
\end{table*}

\section{Method: reproducing the host galaxy photometry by means of
 synthetic and observed SED templates}
\label{method}

The  fit of  the SEDs  have  been carried  out using  Hyperz\footnote{
  http://webast.ast.obs-mip.fr/hyperz/}     (Bolzonella     et     al.
  \cite{Bolz00}).    Eight   synthetic   spectral  types   were   used
  representing Starburst galaxies  (Stb), Ellipticals (E), Lenticulars
  (S0), Spirals (Sa, Sb, Sc  and Sd) and Irregular galaxies (Im).  The
  time evolution of the SFR for  all galaxy types is represented by an
  exponential model,  i.e.  SFR $\propto  \exp(-t/\tau)$, where $\tau$
  is  the SFR  time scale.   Each galaxy  type has  a value  of $\tau$
  assigned.  The  SFR of Stb is  simulated by an  exponential decay in
  the limit when $\tau \rightarrow 0$, in other words an instantaneous
  SFR given  by a delta function.   The early type  galaxy spectra (E,
  S0) are  represented by values of  $\tau$ between 1 and  2 Gyr.  The
  Spiral galaxies (Sa, Sb, Sc  and Sd) have $\tau$ values ranging from
  3 to 30  Gyr.  The SFR of Im galaxies are  represented by a constant
  SFR ($\tau \rightarrow \infty$).

Once the population of stars is generated following the time evolution
given by the assigned SFR, the  mass of the newly formed population is
distributed in stars following an assumed Initial Mass Function (IMF).
Three  IMFs have  been  considered: Miller  \& Scalo  (\cite{Mill79}),
Salpeter (\cite{Salp55}), and  Scalo (\cite{Scal86}).  These IMFs will
be abbreviated  hereafter as MiSc79, Sa55 and  Sc86, respectively.  In
Sect. \ref{impact}  we discuss the impact  of the assumed  IMFs in the
determination of the photometric redshift.

The newly formed stars evolve  depending on their mass and metallicity
following stellar tracks. In  each evolutionary stage the contribution
of all  the individual star  spectra are added yielding  an integrated
galaxy SED which evolves with time.  For each galaxy type the evolving
SEDs can be tabulated and stored creating the so called SED libraries.
Bruzual \&  Charlot (\cite{Bruz93}) have derived a  SED library called
GISSEL98  (Galaxy  Isochrone  Synthesis Spectral  Evolution  Library),
which is the base of our SED fits.

In  addition  to  the  above mentioned  evolutionary  templates,  four
averaged  templates (constructed  grouping  the SEDs  of the  observed
local  galaxies)  from Coleman,  Wu  \&  Weedman (\cite{Cole80})  were
considered (hereafter  named as CWW).  These  extra spectral templates
work  as  a backup  of  the evolutionary  fitting  SEDs,  and give  an
approximate hint of the galaxy type when synthetic SED fits fail.  The
observed CWW templates can be grouped in four sets: early galaxy types
(E/S0), Sbc, Scd and Im.

Furthermore, the impact of  considering different extinction laws has  been
studied.  Four  extinction  laws  have  been taken   into account for   the
determination  of the photometric redshift  and the  physical conditions of
the GRB~000210 host galaxy, namely Calzetti et al.  (\cite{Calz00}), Seaton
(\cite{Seat79}),   Fitzpatrick (\cite{Fitz86}),   and   Pr\'evot  et   al.  
(\cite{Prev84}).  Each    of these laws determine     the dependence of the
extinction on the photon frequency and are the result of different physical
conditions    in  the  interstellar   space  in  the   hosts.  Thus, Seaton
(\cite{Seat79}), Fitzpatrick    (\cite{Fitz86}),  and  Pr\'evot   et   al.  
(\cite{Prev84}),  are appropriate  to describe  the Milky  Way (MW),  Large
Magellanic  Cloud (LMC)  and  the Small Magellanic  Cloud (SMC)  extinction
laws, respectively.  The Calzetti et al.  (\cite{Calz00}) extinction law is
suitable for starburst regions.  In  Sect.  \ref{effect} the effect of  the
adopted extinction law on the inferred  host galaxy photometric redshift is
discussed.   In the  SED fits  a  solar  metallicity ($Z=  Z_{\odot} \simeq
0.02$, being  $Z$ the mass fraction  of heavy elements  in the interstellar
gas) have been assumed.

\begin{figure}[t]
\begin{center}
  {\includegraphics[width=\hsize]{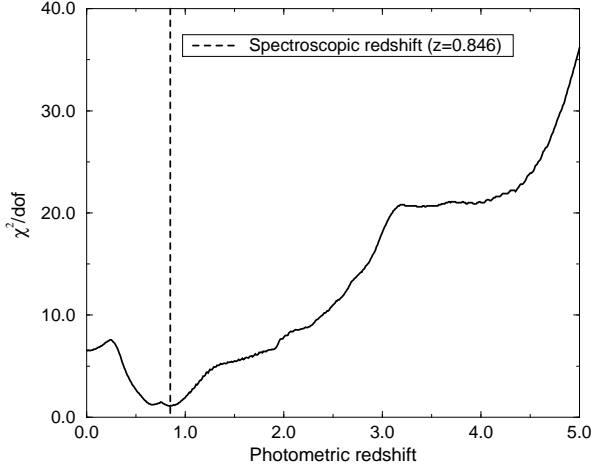}}
  \caption{\label{chi2} The  evolution of the  fitted SED $\chi^2/$dof
  as a function of the  photometric redshift. The dotted vertical line
  indicates  the  spectroscopic  redshift  proposed  by  Piro  et  al.
  (\cite{Piro02}).    As  shown  the   minimum  of   $\chi^2/$dof  (at
  $z=0.842$) is consistent with the spectroscopic redshift.  }
\end{center}
\end{figure}

We varied the GRB~000210 host  galaxy redshift between $z=0$ and $z=5$
with a redshift  step of $\Delta z=0.01$.  The  host galaxy extinction
was  ranged in  a $A_{\rm  V}= 0-5$  interval with  a step  of $\Delta
A_{\rm V}=0.005$  mag.  Table \ref{table3} shows  several inferred fit
parameters for the assumed extinction law and IMFs: the fit confidence
level ($\chi^2/$dof), the photometric redshift $z$ (and the associated
asymmetric  uncertainties),  the best  fitted  template, the  dominant
stellar age, the extinction $A_{\rm V}$, the absolute B-band magnitude
($M_B$), and the host galaxy luminosity (in units of $L^{\star}$).  As
it is shown in Table  \ref{table3} the resolution of our template grid
is not able to make a distinction between most of the properties (Age,
$A_{\rm  V}$, $M_B$, $L/L^{\star}$)  derived for  the Sa55  and MiSc79
IMFs.   Fig.~\ref{chi2}  shows  the  evolution of  $\chi^2/$dof  as  a
function of the best fitted SED  redshift, when a Sa55 IMF is assumed.
The  fit  to the  UBVRIZJsHK-band  photometric  points  shows a  clear
minimum  around $z  \sim 0.85$  and has  no other  acceptable redshift
solutions.

\section{Study of the host galaxy environment}
\label{env}
At  present it  is unknown  if  GRB host  galaxies preferentially  are
located in dense environments, or  if there is any correlation between
the local density  of galaxies and the presence of a  GRB.  So far the
two z=2.04 bursts,  GRB~000301C and GRB~000926, are the  only ones for
which the  environment of the host  galaxy has been  studied (Fynbo et
al.  \cite{Fynb02}).  In both of  these fields a number of galaxies at
the same redshift were detected,  but the lack of blank fields studied
to similar depth  prevented those authors to conclude  if the GRB host
fields were  overdense.  The photometric redshifts of  the galaxies in
the  GRB~000210  field  provide  the  opportunity to  look  for  other
galaxies  in its  environment.   The same  calibration and  photometry
software used to  obtain the host galaxy magnitude  was applied to the
rest of the objects in the field.

\begin{figure}[t]
\begin{center}
  {\includegraphics[width=\hsize]{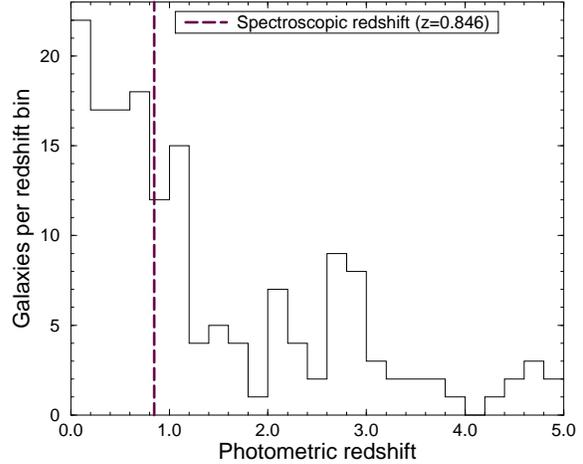}}
\caption{\label{hist}  The  plot shows  the  redshift distribution  of
galaxies  in  a area  of  $6^{\prime}  \times  6^{\prime}$ around  the
GRB~000210 host.  The sample is made up of the 169 galaxies whose SEDs
were fitted with $\chi^2/dof <  2$. The vertical dashed line shows the
spectroscopic redshift.  As shown there is no special concentration of
galaxies  with  redshifts  similar   to  the  host  galaxy.   For  the
construction  of  the  histogram  we  assumed a  MiSc79  IMF  and  the
extinction law given by  Calzetti et al.  (\cite{Calz00}).  Other IMFs
and extinction laws yield similar results.}
\end{center}
\end{figure}

We consider that an object is suitable for redshift determination when
it is detected at least in  four bands.  Objects detected in less than
four  filters  were rejected  due  to  the  large uncertainty  in  the
determination  of their  redshifts.   The considered  region covers  a
$6^{\prime} \times  6^{\prime}$ ~area around the host  galaxy.  At the
redshift of the host galaxy  ($z=0.8463$) this corresponds to $\sim 3$
Mpc  $\times 3$  Mpc.   The  SEDs used  to  determine the  photometric
redshifts of the field objects  consist of 8 synthetic templates (Stb,
E, S0, Sa,  Sb, Sc, Sd, Im)  based on a MiSc79 IMF  and the extinction
law given by  Calzetti et al. (\cite{Calz00}).  As in  the case of the
host SED, four additional observed spectra from CWW were considered.

Among the 169 galaxies of the field with acceptable fits ($\chi^2/$dof
$< 2$)  we considered the  ones with photometric  redshifts compatible
(within $1 \sigma$) with a $\Delta  z = \pm 0.1$ redshift range around
the host  galaxy spectroscopic redshift. In  Fig.~\ref{3.6ESOV} a deep
V-band image around the host  galaxy is displayed.  The FOV covered by
the image  is $5^{\prime}  \times 5^{\prime}$.  The  circles represent
the galaxies having photometric  redshifts consistent with $0.7463 < z
< 0.9463$.  The  radius of each circle is  proportional to the inverse
of the  absolute B-band magnitude ($1/|M_B|$) of  the galaxy contained
inside.  From the distribution of the circles on the image it is clear
that there  is no  obvious concentration of  galaxies around  the host
(indicated by the tick marks).  The lack of clustering around the host
galaxy can  also be visualised in Fig.~\ref{hist},  where the redshift
distribution   of   the   galaxies   in   the   field   are   plotted.
Fig.~\ref{hist}  shows that  there  is  no spike  of  galaxies at  the
redshift of  the host.  The  same study was performed  considering the
200  galaxies  with  SEDs  fitted  having $\chi^2$/dof  $<  3$,  again
yielding  no apparent concentration  of objects  around the  host. The
exercise was  repeated using several extinction laws  and IMFs, giving
similar results.

\begin{figure*}[t]
\begin{center}
  {\includegraphics[width=16cm]{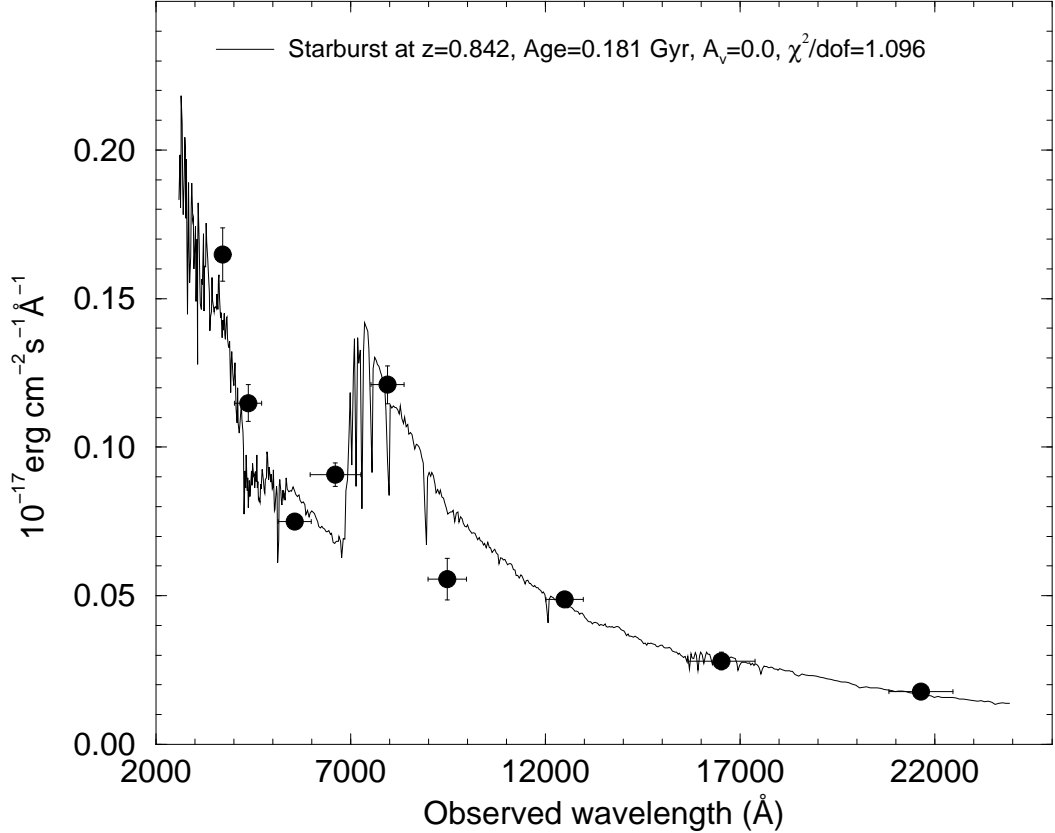}}
\caption{\label{210sed}  The  points show  the  measured  flux in  the
UBVRIZJsHKs  bands   for  the  GRB~000210  host,   once  the  Galactic
dereddening is introduced (Schlegel et al.  \cite{Sche98}).  The solid
curve  represent  the  best  SED  fitted  to  the  photometric  points
($\chi^2$/dof  =  $1.096$),  corresponding  to a  starburst  synthetic
template at a redshift of $z  = 0.842$ generated with a Sa55 IMF.  The
derived value  of the starburst  age corresponds to $0.181$  Gyr.  The
fit is  consistent with a very  low extinction, ($A_{\rm  V} \sim 0$).
The extinction law used to construct  the plot is given by Calzetti et
al.   (\cite{Calz00}).  The SED  shows a  prominent $\sim  4000 \times
(1+z)$~\AA  ~break, bracketed  between  the R  and I-band  photometric
points,  typical of  early  galaxy  types (Stb,  E,  S0) with  stellar
population ages $>10^8$ yr.}
\end{center}
\end{figure*}

 We  have  calculated the  neighbour  detectability  threshold of  our
 images; in other words, the  minimum luminosity of a neighbour galaxy
 for which our data  allows a photometric redshift determination. With
 this purpose  the GRB~000210  host galaxy SED  has been  dimmed until
 detecting  it only  in  seven bands  (UBVRJsHKs)  above the  limiting
 magnitudes given in Table \ref{table1}.  We consider that the minimum
 number of bands to have an acceptable redshift determination is four.
 So,  if  the host  galaxy  SED  was dimmed  in  all  filters by  1.16
 magnitudes (see the limiting magnitudes of Table \ref{table1} and the
 host  galaxy magnitudes of  Table \ref{table2}),  it would  have been
 still  detected in  seven  bands and  a  secure photometric  redshift
 determination would have been still possible. If the SED is dimmed by
 more than $\sim$1.29 mag then  the host would have been detected only
 in RJ  (and may  be marginally in  HK), so no  redshift determination
 would have been possible.
 
 The  absolute   B-band  magnitude  of  a  host   galaxy-like  SED  at
$z=0.8463$,  1.16 mag  shallower, is  $M_B=-19.0$.  So,  a photometric
redshift of a  neighbour galaxy (with a SED similar  to the host) with
$M_B  >  -19.0$  would   have  been  indeterminable.   This  magnitude
corresponds  to  a   Luminosity  of  $L=0.23  L^{\star}$  (considering
$M^{\star}_{B}=-20.6$, following  Schechter \cite{Sche76}).  The value
deduced   for   $L$   based   on   Lilly   et   al.    (\cite{Lill95})
($M^{\star}_B=21.33$,    discussed   in    Sect.    \ref{subluminous})
corresponds  to  $L=0.12   L^{\star}$.   Thus  $M_B=-19.0$  implies  a
luminosity ranging  from $0.12$ to  $0.23 L^{\star}$ depending  on the
adopted $M^{\star}_{B}$ value.  Therefore we consider $L=0.18 \pm 0.06
L^{\star}$ as  an indication  of the limiting  luminosity of  our host
environment study.   This procedure assumes  that the galaxies  in the
host environment have similar SEDs,  thus their SEDs can be reproduced
by dimming  the GRB~000210 host galaxy  SED by the same  factor in all
bands.

\section{Discussion}
\label{discussion}

\subsection{The impact of the assumed IMF on the photometric redshift}
\label{impact}

As  shown   in  Table  \ref{table3}  the   spectroscopic  redshift  is
consistent (within the 99\%  percentile error range) with the inferred
three photometric redshifts. Thus the effect of the assumed IMF is not
crucial to  confirm the spectroscopic redshift of  the GRB~000210 host
galaxy.  However,  among the  assumed three IMFs  the MiSc79  and Sa55
IMFs  are the only  ones providing  a photometric  redshift consistent
within 1$\sigma$  with the spectroscopic redshift.   Thus, we consider
the Sc86 IMF  as the less appropriate one  to describe the predominant
stellar population of the GRB~000210 host galaxy.

 According to Bolzonella et  al. (\cite{Bolz00}) the Sa55 IMF produces
 an excess  of bright blue stars  yielding an UV flux  excess.  On the
 other hand the  Sc86 IMF generates an excessive  number of solar mass
 stars, making  the spectrum too  red to reproduce the  observed SEDs.
 Intensive photometric redshift studies have shown that the MiSc79 IMF
 is  a good  compromise  between both  tendencies  (Bolzonella et  al.
 \cite{Bolz02}).

 In the  particular case of the  blue SED of the  GRB~000210 host, the
 potential excess  of massive stars  given by the  Sa55 IMF is  not an
 inconvenient at  all.  The  prominent UV flux  predicted by  Sa55 can
 easily reproduce the blue part of the observed SED.  On the contrary,
 the excess of solar  mass stars given by the Sc86 IMF  is not able to
 reproduce  the blue  SED  part  unless the  host  galaxy redshift  is
 slightly accommodated.   Thus, the expected impact of  the three IMFs
 (Bolzonella  et   al.   \cite{Bolz02})  is  in   agreement  with  the
 photometric redshifts displayed in Table \ref{table3}.

\subsection{The effect of the adopted extinction law on the photometric 
redshift}
\label{effect}

The  host  galaxy  restframe  SED  flux density  (in  a  $F_{\lambda}$
representation as the one of Fig. \ref{210sed}) increases from 3000 to
2000 \AA ~(corresponding  to the observed SED between the  U and the V
band).  The detection of this ionising UV continuum implies a very low
extinction in the host.  Given the low extinction derived for the host
(see  the values of  $A_{\rm V}$  in Table~\ref{table3})  the inferred
photometric   redshift  is  basically   independent  of   the  adopted
extinction  law  for  the   three  IMFs.   The  results  displayed  in
Table~\ref{table3} for  Sa55 and MiSc79  IMFs remain unchanged  if the
Calzetti et al.  (\cite{Calz00}) extinction law is replaced by another
reddening  law,   as  the   ones  given  by   Seaton  (\cite{Seat79}),
Fitzpatrick (\cite{Fitz86}), or  Pr\'evot et al.  (\cite{Prev84}). The
values of the photometric redshift derived assuming a Sc86 IMF changes
slightly from $z=0.757$ to $z=0.783$, depending on the extinction law.

Therefore, in the particular case of the GRB~000210 host galaxy, the impact
of the adopted extinction  law on the  inferred redshift is  negligible and
has  to be  considered  as a second  order  parameter in  comparison to the
assumed IMF.

\subsection{Is the GRB~000210 host a subluminous  galaxy?}
\label{subluminous}  

A subluminous galaxy  is determined for having a  luminosity below the
 knee  of  the luminosity  function  given  by $L^{\star}$  (Schechter
 \cite{Sche76}).   The characteristic  luminosity  $L^{\star}$ can  be
 associated to  a characteristic AB-system  B-band absolute magnitude,
 $M^{\star}_B(AB)$, which ranges from  $-20.8$ to $-23.0$ depending on
 the rest-frame  colour of the  galaxy (Lilly et  al.  \cite{Lill95}).
 In  a  more  simplified  approximation to  the  luminosity  function,
 Schechter    (\cite{Sche76})    reports    an   unique    value    of
 $M^{\star}_B=-20.6$ (in the Vega system) for all galaxy types.

The restframe  (U$-$V) colour of  the host galaxy  is (U$-$V)$=-0.54$,
which in the AB system corresponds to (U$-$V)$_{\rm AB}=0.15$ (see the
AB  offsets given  in Table  \ref{table2}).  According  to Table  1 of
Lilly et al. (\cite{Lill95})  this (U$-$V)$_{\rm AB}$ colour implies a
value of  $M^{\star}_B(AB)=-21.40$ (given  for a cosmology  defined by
$\Omega_{\Lambda} =  0 $, $\Omega_{M} =  1$ and $H_0=  50$ km s$^{-1}$
Mpc$^{-1}$)  for the  redshift bin  corresponding to  the  host.  This
B-band AB-system magnitude corresponds  to a B-band absolute magnitude
of $M^{\star}_B=-21.33$ in the Vega system (see Table~\ref{table2}).

The absolute B-band  magnitude of the host galaxy  for MiSc79 and Sa55
IMFs ($M^{\star}=-20.16$ see Table \ref{table3}), when rescaled to the
cosmology  used  by  Lilly   et  al  (\cite{Lill95}),  corresponds  to
$M^{\star}   =  -20.18$.    Given   that  $M^{\star}_B=-21.33$,   then
$L=0.35L^{\star}$.  The corresponding value  of $L$ derived for a Sc86
IMF is 0.27 $L^{\star}$ (see last column in Table~\ref{table3}).

The values of $L$, obtained using Schechter (\cite{Sche76}), basically
double (see  the eighth  column in Table~\ref{table3})  those obtained
when   Lilly  et   al  (\cite{Lill95})   is   considered.   Therefore,
considering an averaged  value of $L = 0.5 \pm  0.2 L^{\star}$ for the
host, we conclude  that the host is very  likely a subluminous galaxy.
This  luminosity value  is consistent  with  the one  ($L \approx  0.5
L^{\star}$) derived by Piro et al. (\cite{Piro02}).


\subsection{Estimation of the star formation rate}

The  redshifted  spectra  of  the  GRB~000210  host  galaxy  have  the
restframe  UV  continuum  in  the  observed  optical  range.   The  UV
continuum emission with ongoing star formation is dominated by bright,
short-lived,  main-sequence O  and  B stars.   According to  Kennicutt
(\cite{Kenn98}), for a Sa55 IMF  (consistent with our host galaxy SED,
see Table 3), the SFR in a galaxy is directly proportional to the rest
frame  UV luminosity;  SFR$_{\rm UV}  = 1.4  \times 10^{-28}~L_{\nu}$,
where $L_{\nu}$ indicates the emitted energy per unit frequency around
2800~\AA, measured  in ergs s$^{-1}$ Hz$^{-1}$.   SFR$_{\rm UV}$ gives
the amount of  stellar mass (measured in solar  masses) created in the
host galaxy in a restframe year.   The method of deriving the SFR from
the UV continuum  flux (named SFR$_{\rm UV}$ in  the present paper) is
one of  several diagnostic methods  used in the literature  to measure
SFRs in  galaxies (see  Kennicutt (\cite{Kenn98}) for  a comprehensive
review). Obviously,  if there  is dust-enshrouded star  formation then
this UV-based  method will  only provide a  lower limit to  the actual
SFR.

At $z=0.8463$ the 2800~\AA ~region  is redshifted to 5169.6~\AA, so it
is bracketed  between the  B and  V bands.  Assuming  a power  law SED
stretch between both bands, a  flux density of 0.70 $\pm$ 0.07 $\mu$Jy
is  estimated  at 5169.6~\AA.   This  flux  density  corresponds to  a
restframe  2800~\AA~ luminosity  of $L_{\nu}  = 1.53  \pm  0.15 \times
10^{28}$ ergs s$^{-1}$ Hz$^{-1}$, and therefore to a SFR$_{\rm UV}$ of
$2.1 \pm 0.2 M_{\odot}$ yr$^{-1}$.  The SFR$_{\rm UV}$ derived to date
for GRB  host galaxies range from  1 to 55  $M_{\odot}$ yr$^{-1}$ (see
Berger et  al.  \cite{Berg02}, Table~3).   Thus the SFR$_{\rm  UV}$ of
the GRB~000210 host  galaxy is in the low end  of the distribution for
the   studied  hosts.    The  SFR$_{\rm   UV}$  per   unit  luminosity
(considering $L \sim 0.35 L^{\star}$  based on the Sa55 IMF results of
Table \ref{table3}) is similar to that of other host galaxies.

 As detailed  in Kennicutt  (\cite{Kenn98}) the above  given SFR$_{\rm
UV}$  estimate is  more  adequate for  galaxies  with continuous  star
formation (over time  scales of $10^8$ years or  longer), and provides
an  upper  limit  for  younger  populations such  as  young  starburst
galaxies  with ages  below $10^8$  years.  For  the  estimated stellar
population age of the host galaxy (0.181 Gyr, see Table \ref{table3}),
we  consider  that  the  SFR$_{\rm  UV}$  expression  gives  still  an
acceptable  approximation  to the  actual  SFR$_{\rm UV}$.   Kennicutt
(\cite{Kenn98}) estimates that the internal uncertainty of this method
is $\sim$30\%.   This value is far  from the SFR derived  by Berger et
al.  (\cite{Berg02})  based on the  tentative sub-millimeter detection
of the host galaxy  (SFR$_{\rm smm} \approx 500 M_{\odot}$ yr$^{-1}$).
The apparent  discrepancy between  SFR$_{\rm smm}$ and  SFR$_{\rm UV}$
can not  be explained  by the internal  uncertainties inherent  to the
SFR$_{\rm UV}$ or SFR$_{\rm smm}$ methods.

If the tentative detection of  sub-mm emission from the host galaxy of
GRB~000210  is real,  as opposed  to  noise or  emission from  another
source along the  line of sight, we need to conclude  that the host of
GRB~000210  has two  separate  populations of  massive  stars. One  is
traced  by  the rest  frame  UV/optical light  and  shows  no sign  of
extinction and  the other is completely  obscured by dust  and is only
detectable  at sub-mm  wavelengths.  A  possible way  to  explain this
apparently odd  configuration is if the  host has a  clumpy and opaque
ISM with no  thin absorbers, which is able to  completely hide part of
the massive stellar population,  but does not significantly affect the
UV  flux of  the  not hidden  massive  stars. This  scenario would  be
consistent with the significant  line of sight column density inferred
from  the afterglow  X-ray spectrum  ($N_{\rm H}  = (5  \pm  1) \times
10^{21}$~cm$^{-2}$,  Piro  et  al.   \cite{Piro02}).   It  would  also
naturally  explain  the lack  of  optical  afterglow  emission if  the
progenitor was a member of the enshrouded population.

Based  on the  flux of  the  [\ion{O}{II}] line  and assuming  several
reasonable hypotheses Piro et al.  (\cite{Piro02}) deduced a SFR$_{\rm
[\ion{O}{II}]}$ of  $\sim3 M_{\odot}$ yr$^{-1}$.  Given  the impact of
their assumptions  (they calibrated the GRB~000210 [\ion{O}{II}] flux
relative to the  one of the GRB~970828 host  galaxy) and the intrinsic
scatter  of  the  SFR$_{\rm   UV}$  method  ($\sim$30\%  according  to
Kennicutt \cite{Kenn98}), we consider that our SFR$_{\rm UV}$ estimate
is in agreement with  the SFR$_{\rm [\ion{O}{II}]}$ determined by Piro
et  al.  (\cite{Piro02}).   Thus, the  [\ion{O}{II}] line  and  the UV
continuum originate from the same unextincted blue stellar component.

\subsection{Implications of the fitted SED on the GRB progenitor}

 The fitted SED  assuming a MiSc79 or a Sa55 IMF  is consistent with a
 Stb, independently  of the extinction  law used. The Stb  template is
 characterized by a  value of $\tau \rightarrow 0$, so  the SFR can be
 expressed by a delta function.   In this scenario, the star formation
 is  instantaneous, and occurs  at the  same time  for all  the stars,
 independently of  their masses.  Thus  all the stars should  have the
 same age.  The local birth places  in a host galaxy (even in the same
 star forming region) show different physical conditions and, besides,
 they would be causally separated from each other, so an instantaneous
 star formation  is physically inviable.   Therefore, this description
 should be considered only  as an idealisation of a quasi-simultaneous
 starburst  episode occurred  around 0.181  Gyr ago  (measured  in the
 restframe) in the host galaxy.
 
 Several alternatives are possible to explain a GRB progenitor with an
 age of $\sim$0.181 Gyr.  The  first alternative would be a progenitor
 made up of a binary merging  system. The life time of such systems is
 $\sim$ 0.1 --  1 Gyr, i.e.  compatible with  the host galaxy dominant
 stellar   age   (Eichler   et   al.    \cite{Eich89}).    Thus,   the
 $\gamma$-bright (but  optically dark) GRB~000210 would  come from the
 collapse  of  a compact  binary  system.   This interpretation  would
 support a  connection between dark  GRBs and binary  merging systems,
 which would not necessarily invoke  a circumburst dense region and an
 extinction  mechanism  to  explain   the  lack  of  optical  emission
 (Castro-Tirado  et al.  \cite{CastT02}).   However, a  binary merging
 origin shows several problems.  Piro et al. (\cite{Piro02}) derived a
 column density  of $N_{\rm H} =  (5 \pm 1)  \times 10^{21}$ cm$^{-2}$
 along the line  of sight to the burst. It is  not obvious to conceive
 such  binary systems  located within  a  high density  ($N_{\rm H}  >
 10^{21}$ cm$^{-2}$)  region.  Each of the components  of such systems
 is the result  of an asymmetric collapse of  stellar cores, providing
 in the  instant of the explosion  kick off velocities  up to $900$~km
 s$^{-1}$  to   the  newly  formed   compact  object  (Frail   et  al.
 \cite{Frail94};  Nazin \&  Postnov \cite{Nazi97}).   Thus  the binary
 systems tend to be located far  from their birth places, as they have
 $0.1-1$  Gyr to  travel before  the binary  collapse  episode occurs.
 However, Belczynski et al.  (\cite{Belc02}) have recently shown that,
 although far from the star forming regions, the binary systems should
 occur inside the host galaxies.   Besides they find that such systems
 are more numerous than previously thought.

 In principle a  collapsar with an age of $\sim$0.181  Gyr is not easy
 to accommodate. The age of a  8$M_{\odot}$ star when it explodes as a
 type {\rm  II} SN  is $\sim$0.05 Gyr  (see for instance  Portinari et
 al. \cite{Port98}).  More massive stars, as the progenitors suggested
 in  the collapsar scenario  (Paczy\'nski \cite{Pacz98};  MacFadyen \&
 Woosley \cite{MacF99}), have even shorter lifetimes.

 The clumpy  ISM scenario  would be able  to reconcile  the difference
 between the age  derived from the SED ($\sim$ 0.181  Gyr) and the age
 expected for  a collapsar (the  lifetime of a $\sim  100$ $M_{\odot}$
 progenitor  is  $\sim  0.003$  Gyr,  according to  Portinari  et  al.
 \cite{Port98}).   In such  scenario  the hidden  population of  young
 stars would be able to generate a collapsar but not contribute to the
 host galaxy SED.

\section{Conclusions}

 We have  presented an intensive UBVRIZJsHKs broad  band photometry of
 the  GRB~000210 host  galaxy which  has allowed  us to  determine its
 photometric   redshift.    The   derived  photometric   redshift   is
 $z=0.842^{+0.054}_{-0.042}$,   in   excellent   agreement  with   the
 spectroscopic  redshift ($z=0.8463\pm  0.0002$) proposed  by  Piro et
 al.\ (\cite{Piro02}).  The inferred redshift is basically independent
 of  the extinction law  and IMF  assumed, although  (at least  in the
 particular case of GRB~000210) the Scalo (\cite{Scal86}) IMF provides
 slightly  worse  results than  Miller  \&  Scalo (\cite{Mill79})  and
 Salpeter  (\cite{Salp55})  IMFs.   The  SED  of the  host  galaxy  is
 consistent with a  starburst template with an age  of $\sim$0.181 Gyr
 and  a  very low  extinction  ($A_{\rm V}  \sim  0$).   Based on  the
 restframe UV  flux a star formation  rate of $2.1  \pm 0.2 M_{\odot}$
 yr$^{-1}$ is estimated.

 The absolute restframe B-band magnitude  of the host ($M_B = -20.16$)
 is consistent with  the distribution of the $M_B$  host galaxy values
 measured to date (see Djorgovski  et al.  \cite{Djor01}, Fig. 2).  We
 determine a value of $L  = 0.5\pm0.2 L^{\star}$ for the luminosity of
 the  host, in  agreement  with the  value  estimated by  Piro et  al.
 (\cite{Piro02}).
 
 We have tried to the explore the role played by galactic interactions
 triggering the  GRB phenomena.  Many  host galaxies observed  to date
 appear as part of complex and interacting systems (GRB~980613, Hjorth
 et   al.   \cite{Hjor02};   GRB~001007,  Castro   Cer\'on  et   al.
 \cite{Cast02}).  According to our study the GRB~000210 host galaxy is
 a subluminous  galaxy with no interacting companions  above $0.18 \pm
 0.06 L^{\star}$.

 The low value  of the extinction obtained in the  SED fit ($A_{\rm V}
 \sim 0$)  makes  difficult   to  explain  the  optical  darkness  of
 GRB~000210 in  terms of  the global host  galaxy dust  extinction. If
 dust  extinction is  the  reason  of the  lack  of optical  afterglow
 emission,  then the  circumburst region  has to  be very  compact and
 localised around  the progenitor.   This hypothesis would  agree with
 observations   carried  out   for  the   optically-faint  GRB~990705.
 Andersen  et al.   (\cite{Ande02}) have  localised the  optically dim
 GRB~990705 (but  NIR bright, see  Masetti et al. \cite{Mase00})  in a
 face-on spiral galaxy.   Thus given the thin disk  of a spiral galaxy
 ($\sim 0.3$  kpc), the  optical extinction of  GRB~990705 can  not be
 attributed to the global ISM in its host.  This clumpy and fragmented
 ISM  would also  explain  the apparent  discrepancy  between our  SFR
 estimate  (derived from  the galaxy  UV  flux) and  the one  recently
 reported   based  on   the  sub-millimeter   range  (Berger   et  al.
 \cite{Berg02}; Barnard et al.  \cite{Barn02}).

 Several progenitor models have been discussed in order to explain the
 inferred stellar  population age and the low  host galaxy extinction.
 Both  the  collapsar  and  the  binary  merging  models  show  severe
 limitations to explain the visible  stellar age and the line of sight
 \ion{H}{I} column density (derived from the afterglow X-ray spectrum)
 respectively.  A solution to this problem would be the existence of a
 younger population of stars (several Myr of age) hidden by the clumpy
 ISM.   Such population  (which would  include the  progenitor massive
 star)  would   not  have   any  impact  in   the  host   galaxy  SED.
 Morphological information  derived by  HST could verify  the proposed
 ISM  clumpy  scenario  present  in   the  host  galaxy  of  the  dark
 GRB~000210.
 
\section*{Acknowledgments}
J.  Gorosabel acknowledges the receipt of a Marie Curie Research Grant from
the European  Commission.   This work was supported  by  the Danish Natural
Science  Research Council  (SNF).   J.M.  Castro  Cer\'on  acknowledges the
receipt of a  FPI doctoral fellowship  from Spain's Ministerio de Ciencia y
Tecnolog\'\i  a.  J.U.    Fynbo  acknowledges support    from the Carlsberg
Foundation. We acknowledge our referee L.  Piro for fruitful comments.  The
observations presented  in  this paper were obtained   under the ESO  Large
Programmes 165.H-0464(E), 165.H-0464(I) and 265.D-5742(C).

\end{document}